\documentclass[
aps,%
12pt,%
final,%
notitlepage,%
oneside,%
onecolumn,%
nobibnotes,%
nofootinbib,
superscriptaddress,%
noshowpacs,%
centertags]%
{revtex4}
\begin{document}
%
\title{New limits on Magnetic Monopoles searches from accelerator and non-accelerator experiments}

\author{\firstname{Michela} \surname{Cozzi}}
\email[]{Michela.Cozzi@bo.infn.it}
\thanks{We acknowledge the cooperation of the members of the SLIM Collaboration. We
thank our collaborators and many colleagues for discussions and advise, in particular
S. Balestra, S. Cecchini, G. Giacomelli and L. Patrizii}
\affiliation{Dipartimento di Fisica dell'Universit\`a di Bologna and INFN Sezione di Bologna,
	Viale B. Pichat 6/2, I-40127, Bologna, Italy \\ \vspace{0.5cm} {\rm Invited talk at V International Conference on Non-Accelerator New Physics \\ Dubna, Russia, June 2005}}

\begin{abstract}
\vspace{0.5cm}
Here the status of the searches for ``classical Dirac'' Magnetic Monopoles (MMs) at accelerators and for GUT MMs in the cosmic radiation is discussed.
We present recent analysis for ``classical Dirac'' monopoles at accelerators and the lowest flux upper limit for Magnetic Monopoles in the mass range 10$^{5}$ - 10$^{12}$ GeV obtained with the SLIM experiment at the Chacaltaya High
Altitude Laboratory (5290 m a.s.l.).
\end{abstract}
\maketitle
\section{Introduction}
The lack of symmetry between electric and magnetic charges is one of the oldest puzzles in physics. Why is it possible to isolate positive and negative electric charges, but not north and south magnetic poles? At the beginning of the 19th century there were discussions concerning the magnetic content of matter and the possible existence of isolated magnetic charges. While the existence of magnetic monopoles is not excluded by classical electromagnetism, the first convincing argument in favor of such particles was made by Dirac in 1931 \cite{Dirac}. Dirac linked the existence of magnetic monopoles with the quantization of electric charge and established the relation between the elementary
electric charge $e$ and a basic magnetic charge $g$: 
\begin{equation}
	\begin{array}{ccc} e\cdot g = \frac{n\hbar c}{2} = n g_{D}, & & n = 1, 2, ... \\
	\end{array}
	\label{Eq.Dirac}
\end{equation}
where n is an unknow integer and $g_{D} = \hbar c/2e = 68.5e$ is the unit Dirac charge. Here we assume that the
elementary electric charge is that of an electron. If free
quarks exist, Eq.~\ref{Eq.Dirac} should be modified by replacing $e$
with $e/3$, which effectively increases $g$ by a factor of 3.
The existence of magnetic charges and of magnetic currents would symmetrize in form Maxwell's equations. The symmetry would not be perfect since $e\neq g$ but the couplings could be energy dependent and could merge in a
common value at high energies \cite{Rujula}.
There was no prediction for the MM mass; a rough estimate, obtained assuming that the classical monopole radius is equal to the classical electron radius yields $m_{M}\approx g^{2}m_{e}/e^{2}\approx n \cdot 4700 m_{e}\approx n \cdot 2.4$~GeV/c$^{2}$. From 1931 searches for ``classical Dirac'' monopoles were carried out at every new high-energy accelerator employing a variety of direct and indirect methods \cite{GG}.

Interest in monopoles revived in the 70's, following the discovery by 't Hooft and Polyakov \cite{thooft} that there exist monopole solutions to the field equations of theories in which a semi-simple unifying gauge group breaks into a U(1) group plus others. If the unification scale is $M_{GUT}$, the monopole mass \cite{GUT} is $m_{M} \geq M_{GUT}/\alpha_{GUT}$, with $\alpha_{GUT}$ the grand-unified coupling constant. A consequence of these theoretical developments is that monopoles are expected to be extremely heavy.
In GUTs with  $M_{GUT}\simeq 10^{14}-10^{15}$ GeV and $\alpha_{GUT}\simeq 0.025$, $m_{M}> 
10^{16}-10^{17}$ GeV. 
This is an enormous
mass: MMs cannot be produced at any man--made accelerator, 
existing or conceivable. They could only be produced in the first instants of 
our  Universe and can be searched for in the penetrating Cosmic Radiation (CR). 
The most stringent direct limits on GUT superheavy MMs have been set by the MACRO experiment \cite{MACRO}. 

Larger MM masses are expected if gravity is brought into the unification picture, and in
some SuperSymmetric models. 

Intermediate mass monopoles (IMMs) may have been produced in later phase transitions
in the Early Universe, in which a semisimple gauge group yields a U(1) group \cite{lightMM}. IMMs with masses $10^{5}\div10^{12}$~GeV may be accelerated to relativistic velocities in the galactic magnetic field, and in several astrophysical sites. It has been speculated that very energetic IMMs could yield the highest energy cosmic rays \cite{IMMs}.

The lowest mass MM should be stable, since magnetic charge is conserved like electric charge. Therefore, the MMs produced in the Early Universe should still exist as cosmic relics, whose kinetic energy has been affected first by the expansion of the Universe and then by their travel through galactic and intergalactic magnetic fields.

Here we shall review the present experimental situation on MM searches with
emphasis on the recent analysis for ``classical Dirac'' monopoles at accelerators and on recent limits for Intermediate Mass MMs searches with the SLIM experiment \cite{SLIM}.

\section{Properties of magnetic monopoles}

The main properties of MMs  are obtained from the Dirac relation. 
\par 
\noindent - If $n$~=1 and  the basic electric charge is that of the electron, then  the {\it basic magnetic charge} is $ g_D =\hbar c/ 2e=137e/2$. The magnetic charge is larger if  $n>1$ and  if the basic electric charge is $e/3$.
\noindent 
\par 
\noindent - In analogy with the fine structure constant, $\alpha =e^{2}/\hbar c\simeq 1/137$, the {\it dimensionless magnetic coupling constant} is $ \alpha_g=g^{2}_{D}/ \hbar c \simeq 34.25$; since it is $>1$ perturbative calculations cannot be used.
\par
\noindent - {\it Energy W acquired in a magnetic field  B}:~$  W = ng_{D} B\ell = n \ 20.5$ keV/G~cm. In a coherent galactic--length ($\ell\simeq 1$ kpc,  $B\simeq 3~\mu$G), the energy gained by a MM with $ g=g_{D}$ is $ W \simeq 1.8\times 10^{11}$ GeV. Classical poles and IMMs in the CR may be accelerated to relativistic velocities. GUT poles should have low velocities, $10^{-4}<\beta<10^{-1}$. 
\par                 
\noindent- {\it MMs may be trapped in  ferromagnetic materials} by an image force, which  could reach  values of $\sim 10$ eV/\AA.
\par
\noindent- The interaction of a MM magnetic charge with a nuclear magnetic dipole could lead to the formation of a M--nucleus bound system. A monopole--proton bound state may be produced via  radiative capture. Monopole--nucleus bound states may exist for nuclei with large gyromagnetic ratios.
 \par
\begin{figure*}[h!]
\setcaptionmargin{5mm} 
\onelinecaptionsfalse 
	\includegraphics[width=0.67\textwidth]{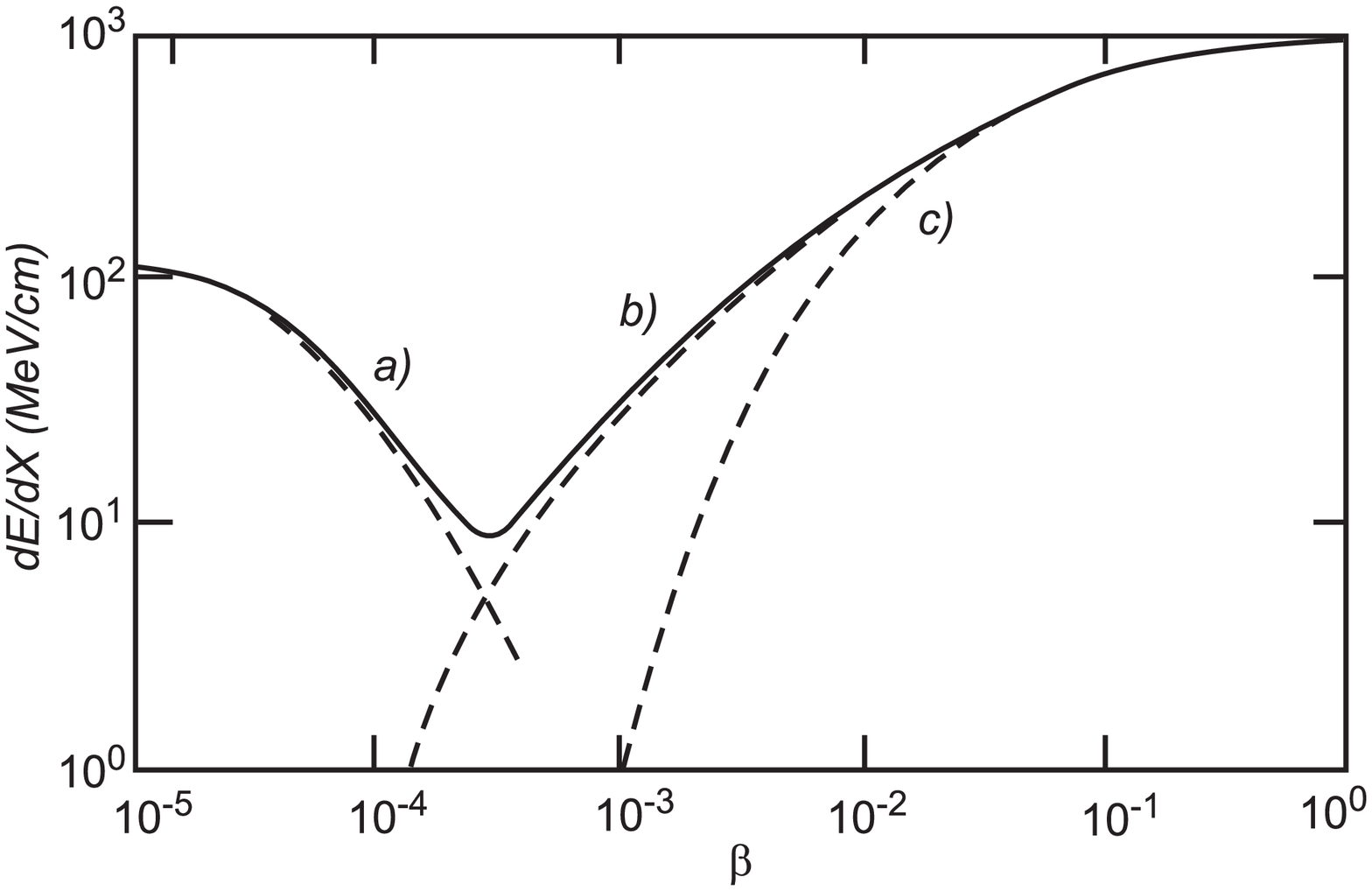}\captionstyle{normal}
	\caption{The energy losses, in MeV/cm, of $g=g_D$ MMs in
liquid hydrogen vs ${ \beta}$. Curve a) corresponds
to elastic monopole--hydrogen atom scattering; curve b) 
to interactions with level crossings; curve c) describes
the ionization energy loss.}
	\label{fig:perdita-di-energia}
\end{figure*}

 \noindent- {\it Energy losses of fast poles.} 
A fast MM with magnetic charge $g_D$ and velocity $v=\beta c$ behaves like an electric charge 
$(ze)_{eq}=g_D\beta$, Fig.\ \ref{fig:perdita-di-energia}.
\par
\noindent - {\it Energy losses of slow poles} ($10^{-4}<\beta<10^{-2}$) may be due to ionization or  excitation of atoms and molecules of the medium (``electronic'' energy loss) or to recoiling atoms or nuclei  (``atomic'' or ``nuclear'' energy loss). Electronic energy loss 
predominates for $\beta>10^{-3}$. 
 \par
\noindent - {\it Energy losses at very low velocities.} 
MMs with   $v<10^{-4}c$ may lose energy in elastic collisions with atoms or with nuclei.  The energy is released to the medium in the form of elastic vibrations and/or infra--red radiation~\cite{derkaoui1}.
\par
Fig.\ \ref{fig:perdita-di-energia} shows  the   energy loss in liquid hydrogen  
of
a $g=g_D$ MM vs  $\beta$~\cite{gg+lp}.\par

\section{Search for ``classical Dirac'' monopoles at accelerators}

By classical monopole we mean a particle without electric charge or hadronic interactions and with magnetic charge $g$ satisfying the Dirac quantization condition (Eq.~\ref{Eq.Dirac}).
Monopole searches at accelerators have predominantly used either induction or ionization methods. 

Induction experiments measure the monopole magnetic charge and are independent of monopole mass and velocity. The method of detection is the search for the induction of a persistent current within a superconducting loop \cite{loop}. Searches for magnetic monopoles using this method have been performed at the $p\bar{p}$ Tevatron collider assuming that produced MMs could stop, be trapped and bound in the matter surrounding the D0 and CDF collision regions \cite{trapped}. Pieces of the detector materials were cut into long thin strips which were
each passed through a superconducting coil coupled to a Superconducting QuantumMechanical
Interference Device (SQUID). Trapped magnetic monopoles in a strip will cause a persistent current to
be induced in the superconducting coil by the magnetic field of the monopole, after complete
passage of the strip through the coil. In contrast, the current due to magnetic dipoles returns
to zero after passage of the strip.

Ionization experiments rely on a magnetic charge producing more ionization than an electrical charge with the same velocity. 
Direct searches for magnetic monopoles using different tracking devices as scintillators, nuclear track detectors (NTDs) or central detectors of complex experiments, were performed at $pp$, $p\bar{p}$ and \epem colliders.
Experiments at Tevatron established cross section limits of $\sim 2\times 10^{-34}$~cm$^{2}$ for MMs with $m_{M} < 850$~GeV  \cite{bertani}, searches LEP excluded masses up to 45~GeV \cite{LEP}.

Indirect searches for classical monopoles have looked for the effects of virtual monopole/anti-monopole loops added to QED Feynman diagrams in $\bar{p}p$ and \epem collisions. 
An indirect search for MMs is the search for {\it multi--$\gamma$ events.} 
Five peculiar photon showers found in emulsion plates exposed to high--altitude CRs, are characterized by an  energetic narrow cone of tens of photons, without any incident charged particle~\cite{multigamma}. 
The total energy of the photons is $\sim 10^{11}$ GeV. The small radial spread of photons  suggested a c.m. $\gamma=(1-\beta^{2})^{-1/2}>10^3$. 
The energies of the photons are too small to have $\pi^o$ decays as their source. 
One possible explanation: a high--energy $\gamma$--ray, with energy  $>10^{12}$ eV, produced a pole--antipole pair, which suffered bremsstrahlung and annihilation producing the final multi--$\gamma$ events. 
Searches for multi-$\gamma$ events were performed at the Tevatron and LEP colliders (Fig.\ \ref{fig:mmclass2}). The D0 experiment searched for  $\gamma$ pairs with high transverse energies; virtual pointlike MMs may rescatter pairs of nearly real photons into the final state via a box monopole diagram; they set a 95\% CL limit of 870 GeV~\cite{abbott}.  At LEP the L3 coll. searched for $Z\rightarrow \gamma\gamma\gamma$ events; no deviation from QED predictions was observed, setting a 95\% CL limit of 510 GeV~\cite{acciarri}. Many authors studied the effects from virtual monopole loops~\cite{Rujula,ginzburg}.
Since the Standard Model $Z^{0}$-boson could couple to monopoles, 
assuming that the coupling between the $Z^{0}$ and a MM pair is larger than
for any lepton pair, the measurement of the $Z^{0}$ decay width provides an indirect limit on MMs production
for $m_{M}<m_{Z}/2$ \cite{Rujula}.

Fig.~\ref{fig:mmclass2} summarizes the cross section limits vs MM mass obtained by direct and indirect experiments (at the Fermilab $\overline p p$ collider, \epem  colliders, the ISR $p p$ collider~\cite{gg+lp}. Most searches are sensitive to poles with magnetic charges $g =n g_{D}$ with $0.5<n<5$.\par

Recently, new limits on Magnetic Monopoles searches have been carried out from OPAL and H1 Collaborations.  
A new direct search for MM pairs produced in the reaction $\epem \rightarrow M\bar{M}(\gamma)$ have been performed by the OPAL collaboration at LEP2. This search is primarily based on the $dE/dx$ in the tracking chamber of the OPAL detector \cite{mio}: due to their large velocities these particles would have high ionization energy losses. This analysis is sensitive to MMs with masses from 45 GeV up to the kinematic limit (about 103 GeV).

The first search for MMs in $e^{+}p$ collisions, at
a centre of mass energy of 300 GeV, was made by the H1 collaboration at HERA \cite{hera}. 
This analysis assumed that heavily ionizing MMs produced in $e^{+}p$ collisions may stop in the beam pipe surrounding the H1 interaction point at HERA. The binding energy of monopoles in the material is expected to be large and so they should remain permanently trapped providing they are stable.  The beam pipe surrounding the interaction region during 1995-1997 (integrated luminosity 60~pb$^{-1}$) was investigated using a SQUID magnetometer with a sensitivity of 0.2~$g_{D}$ to look for stopped magnetic monopoles. No free magnetic charges were observed and charge-dependent upper limits on the cross section for the electro-production of magnetic monopoles have been set.

\section{Search for GUT monopoles in the cosmic radiation}

As  already stated, GUT theories of the electroweak and strong 
interations predict the existence of superheavy MMs 
produced in the Early Universe (EU) when the
GUT gauge group breaks into separate groups, one of which is 
U(1). Assuming that the GUT group is SU(5) (which is excluded by 
proton decay experiments) one should have the following transitions:
\begin{equation}
\footnotesize
    \begin{array}{ccccc}
        {} & 10^{15}\ GeV & {} & 10^{2}\ GeV & {} \\
        SU(5) & \longrightarrow & SU(3)_{C}\times \left[ SU(2)_{L}\times U(1)_{Y}\right] & \longrightarrow & SU(3)_{C}\times U(1)_{EM} \\
       {} & \small10^{-35}s & {} & \small10^{-9}s & {}
    \end{array}
\end{equation}
MMs would be generated as topological point defects in the GUT phase transition, almost
one pole for each causal domain. In the 
standard cosmology this leads to too many poles (the 
monopole problem). Inflation would defer the GUT phase 
transition after large supercooling; in its simplest version 
the number of generated MMs would be very small. However the flux depends critically on several parameters, like the pole mass, the reheating temperature, etc. If the reheating temperature is large enough one would have MMs produced in high energy collisions, like $\epem\rightarrow M\bar{M}$. \par 

\begin{figure*}[t!]
\setcaptionmargin{5mm} 
\onelinecaptionsfalse 
		\includegraphics[width=0.78\textwidth]{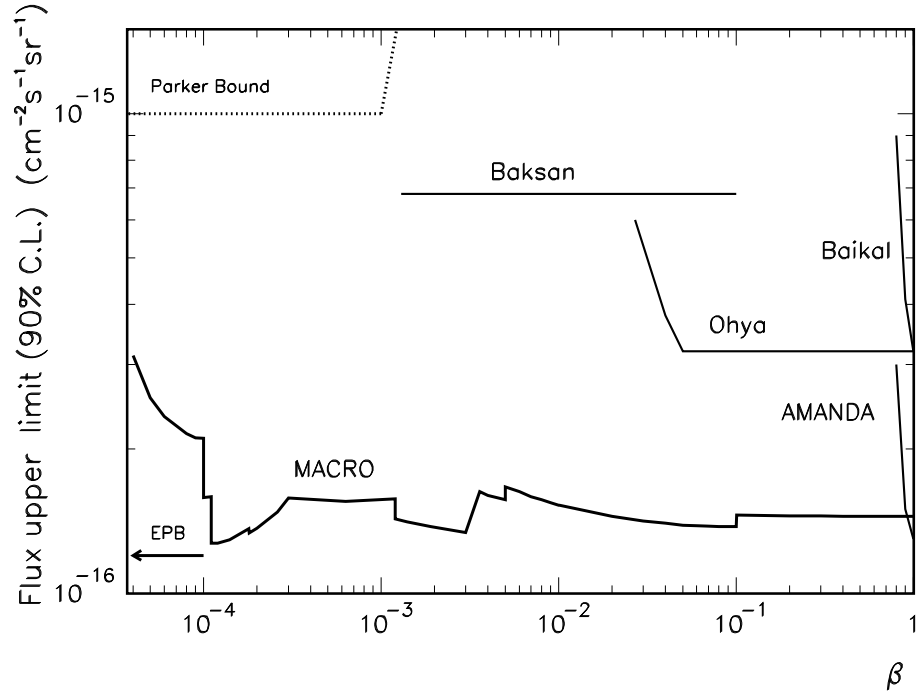}
	\captionstyle{normal}
		\caption{The 90\% CL  MACRO direct upper limits vs $\beta$ for GUT  $g=g_D$ poles in the penetrating CR, and direct limits from other experiments (see text).}
	\label{fig:global2}
\end{figure*}

A flux of cosmic GUT MMs may reach 
the Earth with a 
velocity
spectrum in the range $4 \times 10^{-5} 
<\beta <0.1$,
with possible peaks corresponding to the escape velocities from 
the Earth,
the Sun and the Galaxy.
Searches for such MMs in the  CR 
performed with superconducting induction
devices yielded a combined 90\%~CL limit of
$2 \times 10^{-14}~$cm$^{-2}$~s$^{-1}$~sr$^{-1}$, independent of 
$\beta$~\cite{gg+lp}.
Direct searches were performed above ground and 
underground.
MACRO  performed a search with different   types of 
detectors (liquid scintillators, limited streamer tubes and NTDs)
with an acceptance of  $\sim$ 10,000 m$^2$sr for an isotropic flux.
 No MM was detected;  the  90\% CL flux limits, shown  in
Fig.\ \ref{fig:global2} vs $\beta$  for $g=g_D$, are  at the level of $1.4\times 10^{-16}$~cm$^{-2}$~s$^{-1}$~sr$^{-1}$ for $\beta > 4 \times 10^{-5}$~\cite{MACRO}. The figure shows also the limits from the Ohya~\cite{ohya}, Baksan, Baikal, and AMANDA experiments~\cite{baksan}.  

\section{The SLIM experiment}

The SLIM experiment, which searches for IMMs with NTDs at the Chacaltaya
high altitude lab (5290 m a.s.l.), is sensitive to $g=2g_D$ MMs in  the whole range $4 \times 10^{-5}<\beta <1$ \cite{SLIM}.

The SLIM apparatus at Chacaltaya consists of 440 m$^{2}$ of CR39 and Makrofol nuclear
track detectors. The installation began in March 2000 and was completed in July 2001.
Further 100 m$^{2}$ were installed at Koksil (Himalaya) since 2003.
The detector is organized in modules of 24 cm x 24 cm, each made of 3 layers of
CR39 (1.4 mm thick), 3 layers of polycarbonate (Makrofol, 0.5 mm thick) and of an
aluminium absorber (1 mm thick); each module is sealed in an aluminized plastic bag filled
with dry air.

About 226 m$^{2}$ of CR39 have been etched and analysed, with an average exposure time of 3.6 years. No candidate passed the searching criteria: the 90$\%$ C.L. flux upper limits for fast ($\beta > 0.1$) IMM's coming from above, are at the level of 2.9 10$^{-15}$ cm$^{-2}$ sr$^{-1}$ s$^{-1}$.

\section{Conclusions}

Searches for MMs have shown a remarkable progress since 1931. This is particularly impressive, since no direct evidence of the existence of monopoles has been found.
Even without such evidence, strong theoretical motivations continue to legitimate the experimental program in this field.

\begin{figure*}[h!]
\setcaptionmargin{5mm} 
\onelinecaptionsfalse 
\includegraphics[width=0.9\textwidth]{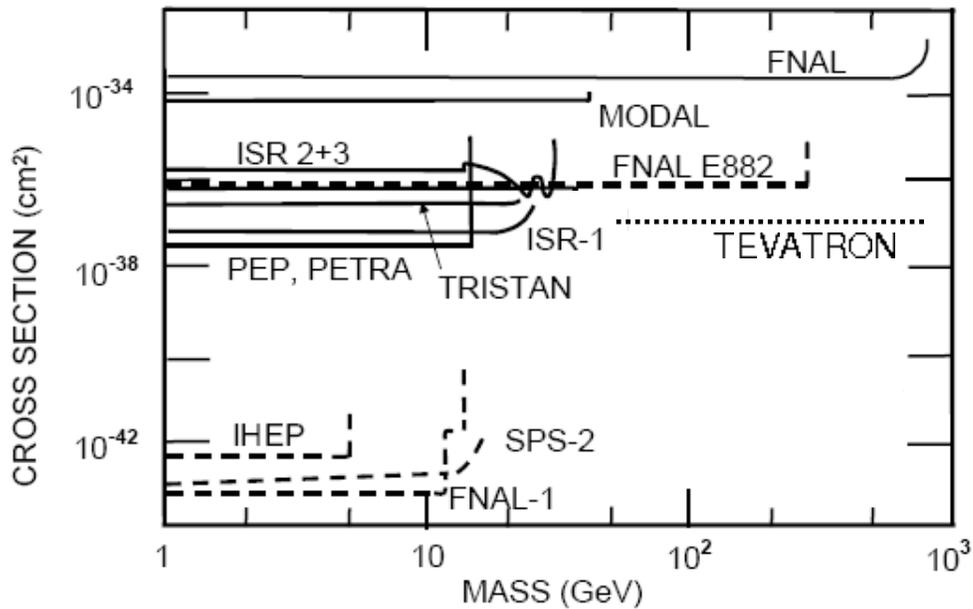}
\captionstyle{normal}
\caption{``Classical Dirac'' MMS cross section upper limits vs MM mass obtained 
from direct and indirect accelerator searches. (solid lines: searches performed with tracking device, dashed lines: other methods.}
\label{fig:mmclass2}
\end{figure*}

Direct and indirect accelerator searches for ``classical Dirac'' MMs
placed limits at the level  $m_M > 850$ GeV with cross section upper values as shown in Fig.~\ref{fig:mmclass2}. Future improvements may come from experiments at Fermilab collider and at the future LHC.

Many searches were performed for heavy GUT MMs in the penetrating cosmic radiation. The 90\% CL flux limits are at $\sim
1.4 \times 10^{-16} $~cm$^{-2}$~s$^{-1}$~sr$^{-1}$ for $\beta \ge 
4 \times 10^{-5}$.
It may be difficult to do much better since one would require refined detectors of
considerably larger areas.

Present limits on Intermediate Mass Monopoles with high $\beta$ are relatively poor. 
By the end of 2006 the SLIM analysis will be completed and the experiment will reach a sensitivity of 10$^{-15}$ cm$^{-2}$ sr$^{-1}$ s$^{-1}$ for $\beta \geq 10^{-2}$ and IMMs with $10^{7} < m_{IMM} < 10^{13}$ GeV. Moreover this search will benefit from the
analysis of further 100 m$^{2}$ of NTDs installed at Koksil (Pakistan).

%
%
\newpage

\end{document}